\documentclass[journal,transmag, hidelinks]{IEEEtran}

\usepackage[left=0.67in,right=0.67in,top=0.66in,bottom=0.67in]{geometry}
\usepackage[T1]{fontenc}
\usepackage[latin9]{inputenc}
\usepackage{amsthm}
\usepackage{amsmath} 
\usepackage{graphicx} 
\usepackage{color} 
\usepackage{cite}
\usepackage{algpseudocode}
\usepackage{algorithm}
\usepackage{amssymb}
\usepackage{subfigure}
\usepackage{esint}
\usepackage{algpseudocode}
\usepackage{epstopdf}
\usepackage{multirow}
\usepackage{makecell}
\usepackage{float}
\usepackage{lipsum}
\usepackage{balance}
\usepackage{tikz}
\usepackage{adjustbox}
\usepackage{bbm}
\usepackage{hyperref}

\newtheorem{definition}{Definition}
\newtheorem{remark}{Remark}

\newtheorem{example}{Example}

\theoremstyle{plain}
\theoremstyle{plain}
\newtheorem{theorem}{Theorem}

\newcommand{\comment}[1]{}

\IEEEoverridecommandlockouts

\begin{document}
%
\title{Non-Binary Constrained Codes for Two-Dimensional \\ Magnetic Recording\vspace{-0.3em}}

\author{
\IEEEauthorblockN{Beyza Dabak,
Ahmed Hareedy,~\IEEEmembership{Member,~IEEE}, and
Robert Calderbank,~\IEEEmembership{Fellow,~IEEE}}

\IEEEauthorblockA{Department of Electrical and Computer Engineering, Duke University, Durham, NC 27708 USA} 
}


\markboth{}%
{}
%



\IEEEtitleabstractindextext{%
\begin{abstract}
The two-dimensional magnetic recording (TDMR) technology promises storage densities of $10$ terabits per square inch. However, when tracks are squeezed together, a bit stored in the two-dimensional (TD) grid suffers inter-symbol interference (ISI) from adjacent bits in the same track, and inter-track interference (ITI) from nearby bits in the adjacent tracks. A bit is highly likely to be read incorrectly if it is isolated in the middle of a $3 \times3$ square; surrounded by its complements, horizontally and vertically. We improve the reliability of TDMR systems by designing two-dimensional constrained codes that prevent these square isolation patterns.
We exploit the way TD read heads operate to design our codes, and we focus on TD read heads that collect signals from three adjacent tracks. We represent the two-dimensional square isolation constraint as a one-dimensional constraint on an alphabet of eight non-binary symbols. We use this new representation to construct a non-binary lexicographically-ordered constrained code where one third of the information bits are unconstrained. Our TD constraint codes are capacity-achieving, and the data protection is achieved with redundancy less than $3\%$ and at modest complexity.
\end{abstract}
\vspace{-0.6em}

\begin{IEEEkeywords}
Constrained codes, two-dimensional magnetic recording, lexicographic ordering, binary to non-binary mapping, data storage.\vspace{-1.1em}
\end{IEEEkeywords}}

\maketitle

\IEEEdisplaynontitleabstractindextext

%
\IEEEpeerreviewmaketitle

\section{Introduction}
%
%
%
%
\IEEEPARstart{O}{ver} the past twenty years, there has been a fierce density competition between magnetic and electronic storage devices. In the last decade, new Flash memory technologies emerged, giving a significant boost in density for electronic storage devices \cite{veeresh_mlc, ahh_qaloco}. Two-dimensional magnetic recording (TDMR) technology is a recent technology that enables magnetic storage devices to stay competitive \cite{wood_tdmr, chan_tdmr, shayan_tdmr, mohsen_tdmr, pituso_tdmr}. In particular, it enables densities of terabits per square inch by squeezing storage tracks together and reducing magnetic isolation. TDMR systems have density approaching $10$ terabits per square inch as reported in \cite{victora_10tbpsi},\cite{hwang_10tb}, whereas one-dimensional magnetic recording can feasibly approach at most $5$ terabits per square inch storage density \cite{seagate}, \cite{mallary_1tb}. The value of TDMR technology is to bridge this difference in storage density. The value of constrained coding is to improve TDMR reliability, and since the redundancy is small, we are able to preserve the dramatic density gains of TDMR.

In data storage, there are certain data patterns that if written to the storage medium, are highly likely to result in errors when read-back. Constrained codes prevent these error-prone patterns from being written. The family of run-length-limited (RLL) codes was introduced in 1970 \cite{tang_bahl}, and was widely used in the 1980's to improve performance of magnetic disks \cite{siegel_mr}. Note that the original presentation of RLL constraints \cite{tang_bahl} described how codewords could be ordered lexicographically. The authors of \cite{ach_fsm} presented a systematic technique to construct constrained codes with rational rates through finite-state machines (FSMs). We refer the reader to \cite{immink_surv} for a useful survey.

We recently introduced lexicographically-ordered constrained codes (LOCO codes) for data storage and data transmission \cite{ahh_loco}, and demonstrated significant density gains in one-dimensional magnetic recording. Then, we introduced asymmetric \cite{ahh_aloco} and $q$-ary asymmetric \cite{ahh_qaloco} LOCO codes to protect the data stored in modern Flash devices. LOCO codes are capacity-achieving, and they offer better rate-complexity trade-offs compared with FSM-based codes. Reconfiguring LOCO codes is as easy as reprogramming an adder \cite{ahh_loco, ahh_aloco, ahh_qaloco}, which helps to manage the device lifecycle.

In TDMR devices, interference comes from the adjacent bits in the same track, which is called inter-symbol interference (ISI), in addition to nearby bits in the adjacent tracks, which is called inter-track interference (ITI) \cite{chan_tdmr, shayan_tdmr}. Equalization, if applied, is assumed to have a target that mimics the channel impulse response \cite{ahh_loco, mohsen_tdmr}. Thus, a data pattern with an isolated bit in the middle of a $3 \times 3$ square grid, surrounded by complementary bits around (left-right, top-bottom, and corners) is error-prone and should be forbidden \cite{mohsen_tdmr}. We call this pattern a \textit{square isolation (SIS) pattern}. Here, we consider the case where a wide head is used to read the bits from three tracks at the same time as shown in \cite{chan_tdmr} and \cite{shayan_tdmr}. Only SIS patterns with their centers aligned with the positions where the wide head will be centered (in middle tracks for each group of non-overlapping three) should be forbidden \cite{chan_tdmr}.

In this paper, we introduce a novel approach to design two-dimensional LOCO (TD-LOCO) codes that forbid the error-prone SIS patterns in order to improve the performance of TDMR devices. We represent each $3$-tuple column of binary bits by a non-binary symbol defined over GF($8$), where GF refers to Galois field, and $q=8$ is its size (order). GF($8$) symbols are partitioned into four groups, each is represented by a GF($4$) symbol such that eliminating one specific pattern of three GF($4$) symbols results in eliminating the SIS patterns. Since we select between two GF($8$) symbols for each GF($4$) symbol, one third of the written bits are effectively unconstrained (one information bit per column). TD-LOCO codes offer modest complexity because of their simple encoding-decoding rule that we derive. TD-LOCO codes are capacity-achieving, reconfigurable, and they protect the data in a TDMR device with less than $3\%$ redundancy. The proposed coding scheme adopting TD-LOCO codes is capacity-approaching with respect to the optimal capacity. More details are discussed in Section~\ref{sec_nbmap}. There is prior work on efficient TD constrained codes \cite{pituso_tdmr, sharov_TCon, halevy_TD, kamabe_TD, kato_TCon, siegel_TCon}, for example, TD-RLL codes, but these codes are either not customized to forbid the error-prone patterns in TDMR systems or do not exploit the properties of modern TDMR systems to reduce redundancy.

The rest of the paper is organized as follows. In Section~\ref{sec_nbmap}, we describe the problem setup and non-binary mapping. In Section~\ref{sec_freebit}, we introduce our novel TD-LOCO codes. In Section~\ref{sec_card}, we enumerate TD-LOCO codewords and derive the encoding-decoding rule. In Section~\ref{sec_rate}, we discuss the rates, give examples, and make comparisons. In Section~\ref{sec_algr}, we introduce the encoding-decoding algorithms and discuss reconfigurability. In Section~\ref{sec_conc}, we conclude the paper.

\section{Problem Setup and Non-Binary Mapping}\label{sec_nbmap}

In this section we describe the TDMR grid, translate the two-dimensional binary constraint to a one-dimensional non-binary constraint, and introduce the proposed base case TD-LOCO codes.

In TDMR, the number of tracks and the number of bits per track depend on properties of the magnetic substrate. A wide read head makes it possible to read multiple tracks at the same time \cite{chan_tdmr}. In this paper, we focus on the scenario in which $3$ of these tracks are read simultaneously. The following SIS patterns should be forbidden because of ISI and ITI effects on the central bit \cite{sharov_TCon}:

\begin{center}
\begin{tikzpicture}
\draw[step=0.5cm] (-1,-1) grid (0.5,0.5);
\draw (2,-1) -- (2,0.5);
\draw[step=0.5cm] (2,-1) grid (3.5,0.5);

\node at (-0.75,+0.25) {0};
\node at (-0.25,+0.25) {0};
\node at (+0.25,+0.25) {0};
\node at (-0.75,-0.25) {0};
\node at (-0.25,-0.25) {1};
\node at (+0.25,-0.25) {0};
\node at (-0.75,-0.75) {0};
\node at (-0.25,-0.75) {0};
\node at (+0.25,-0.75) {0};

\node at (2.25,+0.25) {1};
\node at (2.75,+0.25) {1};
\node at (3.25,+0.25) {1};
\node at (2.25,-0.25) {1};
\node at (2.75,-0.25) {0};
\node at (3.25,-0.25) {1};
\node at (2.25,-0.75) {1};
\node at (2.75,-0.75) {1};
\node at (3.25,-0.75) {1};
\end{tikzpicture}
\end{center}

\begin{remark}\label{remark1}
Here, we assume a TD channel with impulse response (read-head sensitivity) such that the bit at the center of the $3 \times 3$ grid flips only if it is surrounded by complementary bits in all $8$ positions. A more general case of interest is when the $4$ corner bits have limited impact. When the influence of the 4 corner bits is limited, the forbidden patterns take the shape of a plus sign \cite{halevy_TD, mohsen_tdmr}, and we leave the design of LOCO codes for this constraint to our future work.
\end{remark}

Assume there are $N$ tracks (the value of $N$ depends on~the magnetic disk), where $N=3i$ with $i \in \{1, 2, 3, \dots\}$. Denote the tracks by $\{{T}_{0}, {T}_{1}, \dots, {T}_{N-1}\}$.  In the setup we propose, after the wide read head finishes reading $3$ tracks, it continues on reading the following $3$ tracks, e.g., $(T_0, T_1, T_2)$ then $(T_3, T_4, T_5)$. Consequently, the ITI between the tracks at the bottom of the $3$-tuple and the top of the next $3$-tuple is negligible as demonstrated in \cite{chan_tdmr}. This property enables us to design our TD-LOCO codes as non-binary LOCO codes that are used to write the bits on each group of $3$ tracks together as we will illustrate shortly. Bits are written vertically starting from the upper left corner of the TDMR grid.

We suggest a novel approach to solve the discussed two-dimensional binary problem by mapping the problem into a one-dimensional non-binary problem.

First, denote a Galois field (GF) of size $8$ by GF($8$), and let $\beta$ be a primitive element of GF($8$). Therefore,
\begin{equation}
\textup{GF}(8) \triangleq \{0, 1, \beta, \beta^2, \beta^3, \beta^4, \beta^5, \beta^6\}. \nonumber
\end{equation}

Each symbol in GF($8$) corresponds to $3$ bits (standard mapping) that will be written in a grid column. Observe that the SIS forbidden patterns map into:

\begin{center}
\begin{tikzpicture}
\draw[step=0.5cm] (-1,-1) grid (0.5,0.5);
\draw (2,-1) -- (2,0.5);
\draw[step=0.5cm] (2,-1) grid (3.5,0.5);
\draw [thick,dash dot, color=red] (-0.90,-0.93) rectangle (-0.58,0.44);
\draw [thick,dash dot, color=green] (-0.40,-0.93) rectangle (-0.08,0.44);
\draw [thick,dash dot, color=red] (0.10,-0.93) rectangle (0.42,0.44);

\draw [thick,dash dot, color=blue] (2.10,-0.93) rectangle (2.42,0.44);
\draw [thick,dash dot, color=purple] (2.6,-0.93) rectangle (2.92,0.44);
\draw [thick,dash dot, color=blue] (3.10,-0.93) rectangle (3.42,0.44);

\node at (-0.75,+0.25) {0};
\node at (-0.25,+0.25) {0};
\node at (+0.25,+0.25) {0};
\node at (-0.75,-0.25) {0};
\node at (-0.25,-0.25) {1};
\node at (+0.25,-0.25) {0};
\node [label={[red]below:0}] at  (-0.75,-0.75) {0};
\node [label={[green]below:$\beta$}] at (-0.25,-0.75) {0};
\node [label={[red]below:0}] at (+0.25,-0.75) {0};

\node at (2.25,+0.25) {1};
\node at (2.75,+0.25) {1};
\node at (3.25,+0.25) {1};
\node at (2.25,-0.25) {1};
\node at (2.75,-0.25) {0};
\node at (3.25,-0.25) {1};
\node [label={[blue]below:$\beta^6$}] at (2.25,-0.75) {1};
\node [label={[purple]below:$\beta^4$}] at (2.75,-0.75) {1};
\node [label={[blue]below:$\beta^6$}] at (3.25,-0.75) {1};
\end{tikzpicture}
\end{center}

Following this standard bijective mapping between a two-dimensional binary sequence and a one-dimensional non-binary sequence, we now formally define the base case TD-LOCO codes, which are said to be $\mathcal{Q}^8$-constrained.

\begin{definition}\label{def_nbloco}
A base case TD-LOCO code $\mathcal{TC}^8_{m}$ with parameter m is defined by the following properties:
\begin{enumerate}
\item Each codeword $\bold{c}$ in $\mathcal{TC}^8_{m}$ has its symbols from GF(8) and is of length $m$ symbols. 
\item Codewords in $\mathcal{TC}^8_{m}$ are lexicographically ordered.
\item Each codeword $\bold{c}$ in $\mathcal{TC}^8_{m}$ does not contain any of the following two patterns in the set $\mathcal{Q}^8$, where:
\begin{equation}\label{eqn_forbid_original}
\hspace{-1.14em}\mathcal{Q}^8 \triangleq \{0 \beta 0, \beta^{6}\beta^{4}\beta^{6}\}. 
\end{equation}
\item The code $\mathcal{TC}^8_{m}$ contains all codewords satisfying the above three properties.
\end{enumerate}
\end{definition}

Note that, lexicographic ordering means symbol significance reduces from left to right, and codewords are ordered in an ascending manner i.e., $0 < 1 < \beta < \dots < \beta^{6}$ for any symbol.

We calculate the capacity of $\mathcal{Q}^8$-constrained codes, using the following finite-state transition diagram (FSTD):

\begin{figure}[H]
\vspace{-0.0em}
\centering
\includegraphics[trim={1.2in 0in 3.3in 0in},clip,width=3.5in]{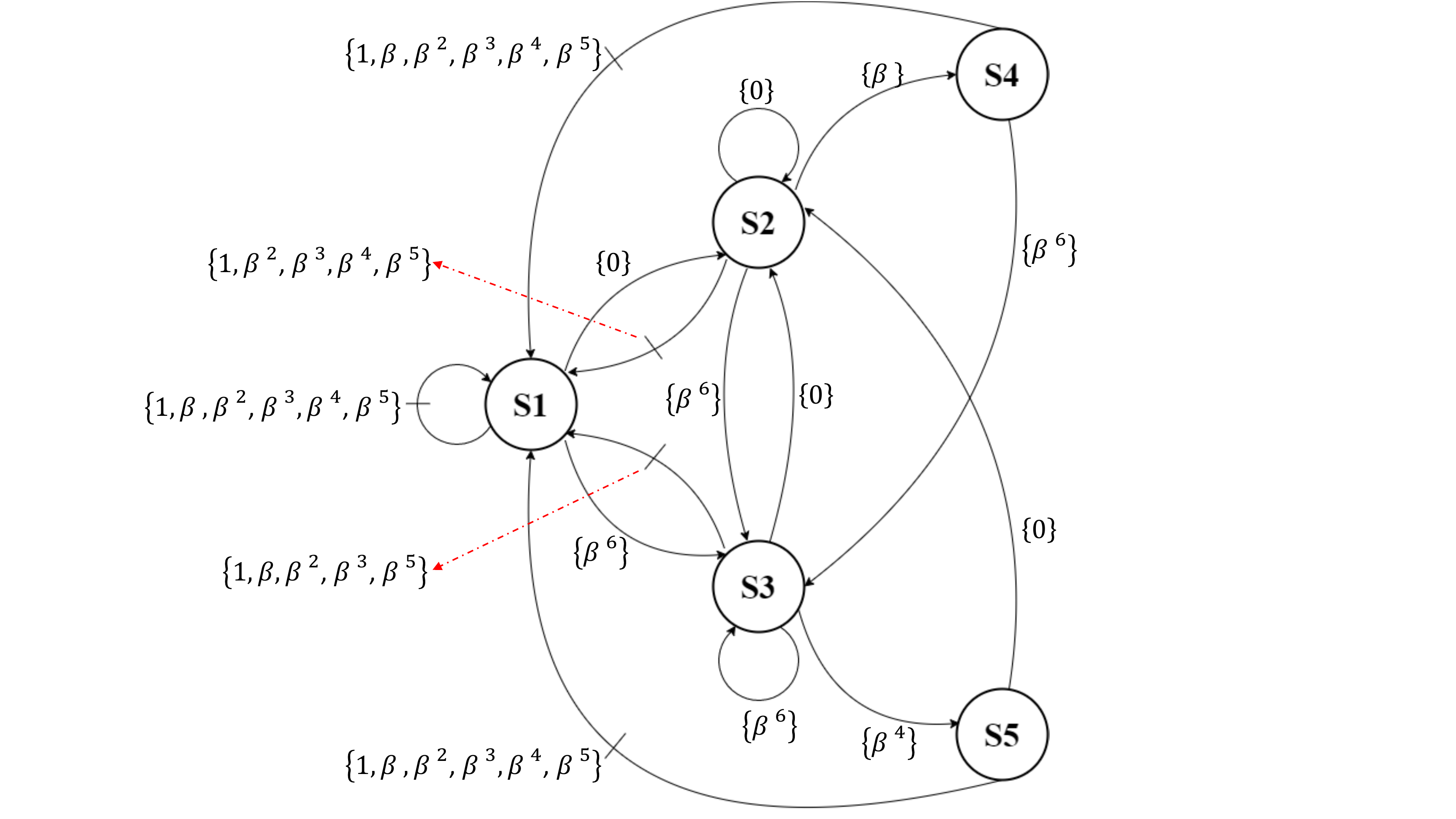}
\caption{FSTD of $\mathcal{Q}^8$-constrained codes}
\label{Fig_fstd1}
\vspace{-0.3em}
\end{figure}

The capacity $C$ of these codes can be obtained from state-transition matrix, which is the adjacency matrix of the FSTD  \cite{siegel_mr}. The state-transition matrix $T$ of this FSTD is given by:

\[
  T=
  \left[ {\begin{array}{ccccc}
   6 & 1 & 1 & 0 & 0\\
   5 & 1 & 1 & 1 & 0\\
   5 & 1 & 1 & 0 & 1\\
   6 & 0 & 1 & 0 & 0\\
   6 & 1 & 0 & 0 & 0\\
  \end{array} } \right].
\]

The capacity $C$ is the binary logarithm of the largest positive eigenvalue $\lambda$ of the matrix $T$ \cite{shannon}, and we have:
\begin{equation}
C =  \log_2 \lambda  = \log_2 7.9690 = 2.9944 .
\end{equation}

This capacity is in information bits per coded symbol, and we defined symbols over GF($8$). Since each GF($8$) symbol represents 3 bits, the normalized capacity (the maximum achievable information rate per binary symbol) is given by:
\begin{equation}\label{capacity_initial}
C^{\textup{n}} = 2.9944/\log_2 8 = 0.9981.  
\end{equation}

Thus, the base case codes have very high capacity. However, simpler codes with lower encoding-decoding complexity can be derived. In the next section, we give up only $0.54\%$ of the available capacity in order to reduce complexity via applying an additional mapping to the proposed codes.

\section{Our Novel TD-LOCO Codes with Additional Information Bit}\label{sec_freebit}

Assume now that we have an alphabet of size $4$. We use symbols from GF($4$) $\triangleq \{0, 1, \alpha, \alpha^2\}$ to represent pairs of symbols from GF($8$). The proposed mapping between GF($8$) symbols and GF($4$) symbols is as follows:
\begin{center}
\begin{tabular}{lcl}
Set 1: $\{\beta,\beta^4\}$ & $\longrightarrow$ & $\{0\}$,\\
Set 2: $\{1,\beta^5\}$ & $\longrightarrow$ & $\{1\}$, \\
Set 3: $\{\beta^2,\beta^3\}$ & $\longrightarrow$ & $\{\alpha\}$,\\
Set 4: $\{0,\beta^6\}$ & $\longrightarrow$ & $\{\alpha^2\}$.\\
\end{tabular}
\end{center}

Recall that in the end, we write binary representation of symbols in GF($8$), i.e., $3$ bits each. Under the new mapping, our constrained code produces symbols in GF($4$), and for each symbol in GF($4$), we select one option out of two from GF($8$) as illustrated in the new mapping. Thus, in the new coding scheme, we have one additional information bit for each grid column because of this selection.

After this mapping, the set $\mathcal{Q}^8$ of forbidden patterns in (\ref{eqn_forbid_original}), which is $\{0 \beta 0, \beta^{6}\beta^{4}\beta^{6}\}$, is mapped into $\{\alpha^2 0 \alpha^2\}$, which is $\mathcal{Q}^4$. The new set subsumes the previous forbidden set, and it contains some additional patterns. The effect of the additional patterns included in the new set appears in capacity. Now, we define the TD-LOCO codes that are $\mathcal{Q}^4$-constrained codes after this mapping.

\begin{definition}\label{def_nbloco_final}
A TD-LOCO code $\mathcal{TC}^4_{m}$ with parameter m is defined by the following properties:
\begin{enumerate}
\item Each codeword $\bold{c}$ in $\mathcal{TC}^4_{m}$ has its symbols from GF($4$) and is of length $m$ symbols. 
\item Codewords in $\mathcal{TC}^4_{m}$ are lexicographically ordered.
\item Each codeword $\bold{c}$ in $\mathcal{TC}^4_{m}$ does not contain the following pattern in the set $\mathcal{Q}^4$, where:
\begin{equation}\label{eqn_forbid_new}
\hspace{-1.14em}\mathcal{Q}^4 \triangleq \{\alpha^2 0 \alpha^2\}. 
\end{equation}
\item The code $\mathcal{TC}^4_{m}$ contains all codewords satisfying the above three properties.
\end{enumerate}
\end{definition}

Our \textbf{overall coding scheme} is such that the constrained code produces symbols in GF(4), each GF(4) symbol corresponds to a pair of GF(8) symbols, and we transmit an additional bit of information through the choice of GF(8) symbol.

We calculate the capacity of the new $\mathcal{Q}^4$-constrained codes, using the FSTD shown below:  

\begin{figure}[H]
\vspace{-0.0em}
\centering
\includegraphics[trim={2.5in 1.8in 2.8in 2.2in},clip,width=3.4in]{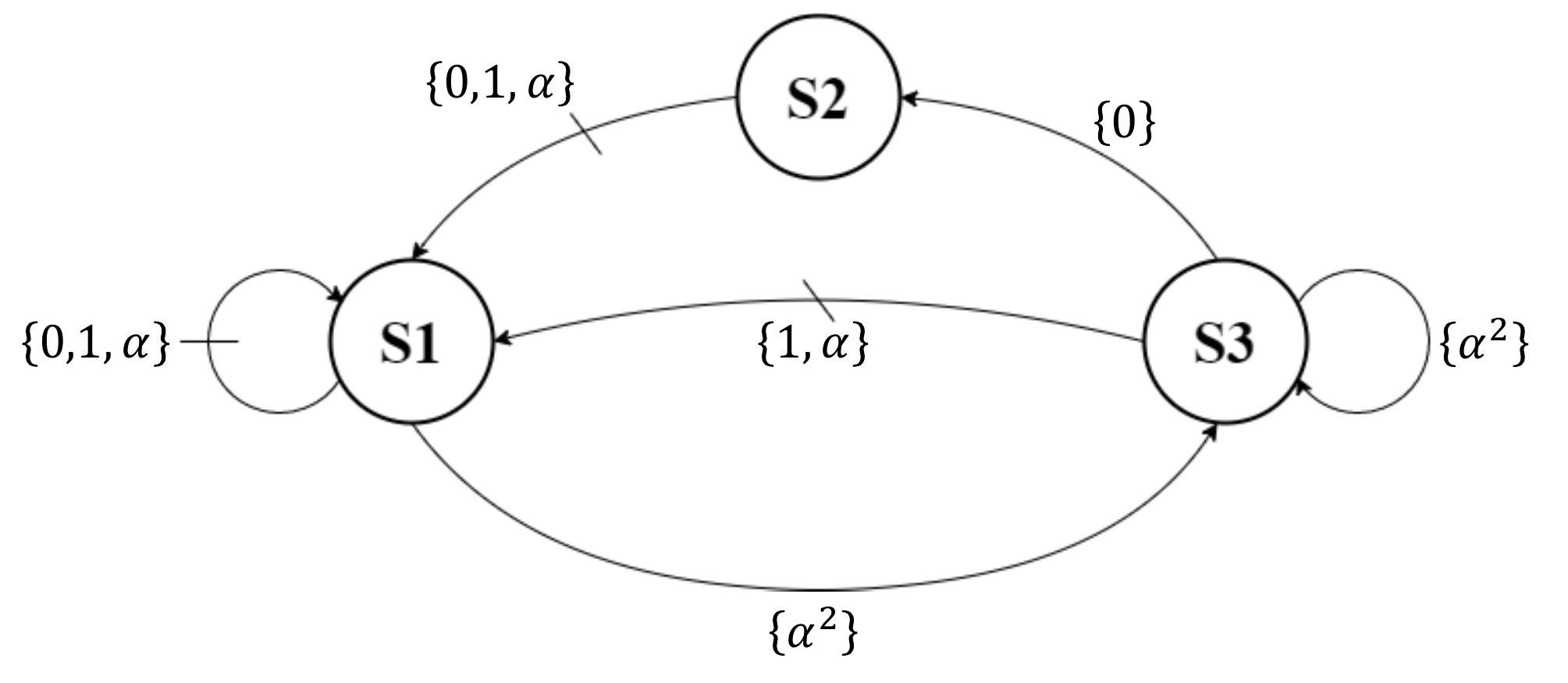}
\caption{FSTD of $\mathcal{Q}^4$-constrained codes}
\label{Fig_fstd1}
\vspace{-0.3em}
\end{figure}

The state-transition matrix derived from the above FSTD is:
\[
  T'=
  \left[ {\begin{array}{ccc}
   3 & 0 & 1 \\
   3 & 0 & 0 \\
   2 & 1 & 1 \\
  \end{array} } \right]
\]

Thus, the capacity of the new constrained codes is:
\begin{equation}
C' =  \log_2 \lambda'  = \log_2 3.9395 = 1.9780.
\end{equation}

Recall that the proposed new code produces symbols in GF($4$). The overall coding scheme is such that one additional information bit is used to select the GF($8$) symbol as discussed above, and each symbol corresponds to $3$ bits. Thus, the normalized capacity of the overall coding scheme is the capacity of the $\mathcal{Q}^4$-constrained codes plus one additional information bit, divided by $3$:
\begin{equation}\label{capacity_new}
C^{\textup{n}}_{\textup{overall}} = \frac{C' + 1}{3} = \frac{2.9780}{3} = 0.9927 .
\end{equation}

The reason why the capacity is less in the proposed scheme is that we add slightly more redundancy. We are forbidding $2^3$ patterns when we group GF($8$) symbols and map them to GF($4$) symbols, whereas in the optimal case, we only need to forbid $2$ patterns as observed in (\ref{eqn_forbid_original}). Nevertheless, the capacity of the proposed overall scheme in (\ref{capacity_new}) is only $0.54\%$ away from the optimal capacity in (\ref{capacity_initial}), whereas the complexity reduction achieved by the proposed scheme is significant. The order of storage and encoding-complexity growth is $O(q-1)$, where $q$ is the GF size, as shown in \cite{ahh_qaloco}. Thus, by operating on GF($4$) instead of GF($8$), we are notably reducing the order of complexity; it is approximately equivalent dividing the order of storage and encoding-complexity by $2.3$. This observation demonstrates the value of the proposed GF($4$) mapping and the addition of a selection bit in the overall scheme. In the next section, we will derive the cardinality and encoding-decoding rule of the TD-LOCO codes.

\section{Cardinality and Encoding-Decoding Rule \\ of TD-LOCO Codes}\label{sec_card}

We now partition the codewords of a TD-LOCO code $\mathcal{TC}^4_{m}$ into $3$ groups based on the symbols they start with from the left, i.e., their left-most symbols (LMSs), to derive the cardinality and encoding-decoding rule:\\
\textbf{Group~1:} Codewords starting with $\delta$ at their LMS, where $\delta \in \{0, 1, \alpha\}$.\\
\textbf{Group~2:} Codewords starting with $\alpha^{2}\theta$ at their LMSs, where $\theta \in \{1, \alpha, \alpha^2\}$.\\
\textbf{Group~3:} Codewords starting with $\alpha^{2} 0 \delta$ at their LMSs.

Observe that given the set of forbidden patterns $\mathcal{Q}^4$ in (\ref{eqn_forbid_new}), all codewords are spanned by the groups. Now, we are ready to enumerate TD-LOCO codewords recursively.

\begin{theorem}\label{thm_card}
The cardinality (size) of a TD-LOCO code $\mathcal{TC}^4_{m}$, denoted by $N(m)$, is given by:
\begin{align}\label{eqn_cardgen}
N(m) &= 4N(m-1) - N(m-2) + 3N(m-3), \textup{ } m \geq 2,
\end{align}
where the defined cardinalities are:
\begin{equation}\label{eqn_cardinit}
N(1) \triangleq 4,  N(0) \triangleq 1, \text{ and } N(-1) \triangleq 1/3.
\end{equation}
\end{theorem}

\begin{IEEEproof}
We use the group structure stated above to prove the recursive formula (\ref{eqn_cardgen}).

\textbf{Group~1:} Each codeword in Group~1 in $\mathcal{TC}^4_{m}$ starts with one of the elements in $\{0, 1, \alpha\}$ from the left, and corresponds to a codeword in $\mathcal{TC}^4_{m-1}$ such that they share the $m-1$ right-most symbols (RMSs). This correspondence is surjective. Since the set $\{0, 1, \alpha\}$ has $3$ elements, the correspondence is $3$ codewords of length $m$ to $1$ codeword of length $m-1$. Thus, the cardinality of Group~1 in $\mathcal{TC}^4_{m}$ is given by:
\begin{equation}\label{eqn_cardg1b}
N_{1}(m) = 3N(m-1).
\end{equation}

\textbf{Group~2:} Each codeword in Group~2 in $\mathcal{TC}^4_{m}$ starts with $\alpha^{2}\theta$ from the left, and corresponds to a codeword in $\mathcal{TC}^4_{m-1}$ that starts with one of the elements in $\{1, \alpha, \alpha^2\}$ from the left such that they share the $m-2$ RMSs. Since the set $\{1, \alpha, \alpha^2\}$ is GF($4$)$\setminus\{0\}$, the codewords in $\mathcal{TC}^4_{m-1}$ that start with $\theta$ from the left are obtained by excluding the codewords in $\mathcal{TC}^4_{m-1}$ that start with $0$ from the left from all the codewords in $\mathcal{TC}^4_{m-1}$. Notice that, codewords in $\mathcal{TC}^4_{m-1}$ that start with $0$ from the left correspond to all codewords in $\mathcal{TC}^4_{m-2}$. Thus, the cardinality of Group~2 in $\mathcal{TC}^4_{m}$ is given by:
\begin{align}\label{eqn_cardg2b}
N_{2}(m) = N(m-1) - N(m-2).
\end{align}

\textbf{Group~3:} Each codeword in Group~3 in $\mathcal{TC}^4_{m}$ starts with $\alpha^{2} 0 \delta$ from the left, and corresponds to a codeword in $\mathcal{TC}^4_{m-2}$ that starts with $\delta$ from the left such that they share the $m-3$ RMSs. These are the codewords of Group 1 in $\mathcal{TC}^4_{m-2}$. This implies that each codeword in Group~3 in $\mathcal{TC}^4_{m}$ corresponds (bijectively) to a codeword in Group 1 in $\mathcal{TC}^4_{m-2}$. Thus, the cardinality of Group~3 in $\mathcal{TC}^4_{m}$ is given by:
\begin{align}\label{eqn_cardg3b}
N_{3}(m) = N_{1}(m-2) = 3N(m-3), 
\end{align}
where the second equality in (\ref{eqn_cardg3b}) is reached aided by (\ref{eqn_cardg1b}) to compute $N_{1}(m-2)$.

Now, the cardinality of $\mathcal{TC}^4_{m}$ is computed using (\ref{eqn_cardg1b}), (\ref{eqn_cardg2b}), and (\ref{eqn_cardg3b}):
\vspace{-0.3em}\begin{align}
N(m) &= \sum_{\ell=1}^3 N_{\ell}(m) \nonumber \\ &= 4N(m-1) - N(m-2) + 3N(m-3), \nonumber
\end{align}
which completes the proof.
\end{IEEEproof}

Now, we derive an encoding-decoding rule, which is a formula that relates the lexicographic index of a TD-LOCO codeword to the codeword itself.

We define a TD-LOCO codeword of length $m$ symbols as $\bold{c} \triangleq c_{m-1} c_{m-2} \dots c_0$ in $\mathcal{TC}^4_{m}$, where $c$ is an element in GF($4$). Define $a \triangleq \mathcal{L}(c)$ as the integer equivalent to symbol $c$, which is given by:
\begin{align}\label{eqn_gflog}
a \triangleq \mathcal{L}(c) \triangleq \left\{\begin{matrix}0, \textup{ } &c = 0,
\\ \textup{gflog}_\alpha(c)+1, \textup{ } &\textup{otherwise},
\end{matrix}\right.
\end{align}
where $\textup{gflog}_\alpha(c)$ returns the power of the GF element $c$ with $\textup{gflog}_\alpha(1)=0$. Thus, the set of integer-equivalents to the elements in GF($4$) becomes $\{0, 1, 2, 3\}$, and the set of forbidden patterns, $\mathcal{Q}^4$, which is defined in (\ref{eqn_forbid_new}), has the integer-equivalent $\{303\}$.

The index of a TD-LOCO codeword $\bold{c}$ in $\mathcal{TC}^4_{m}$ is denoted by $g(m, \bold{c})$, which is sometimes abbreviated to $g(\bold{c})$ for simplicity. For each symbol $c_i$, we define its integer-equivalent $a_i \triangleq \mathcal{L}(c_i)$ as shown above in (\ref{eqn_gflog}), with $c_i \triangleq 0$ and $a_i \triangleq 0$ for $i \geq m$. Our lexicographic index $g(\bold{c})$ is in $\{0, 1, \dots, N(m)-1\}$.

The following theorem introduces the encoding-decoding rule of TD-LOCO codes. Observe that indexing is straightforward for the case of $m=1$.

\begin{theorem}\label{thm_rule}
Consider a TD-LOCO code $\mathcal{TC}^4_{m}$ with $m \geq 2$. Let $\bold{c}$ be a TD-LOCO codeword in $\mathcal{TC}^4_{m}$. The relation between the lexicographic index $g(\bold{c})$ of this codeword and the codeword itself is given by:
\begin{align}\label{eqn_newg}
g(\bold{c}) &= \sum_{i=0}^{m-1} \Big [ a_i N(i) \nonumber \\ &+ \mathbbm{1}(a_{i+1} = 3 \textup{ \& }   a_{i}\neq 0) \hspace{-2.7em}\sum_{\hspace{+2.5em}j \geq 0, \textup{ } i-2j > 0} \hspace{-2.3em} (-1)^{j+1}N(i-2j-1) \Big ],
\end{align}
where $\mathbbm{1}(a_{i+1} = 3 \textup{ \& }   a_{i}\neq 0)$ is an indicator function on whether $c_{i+1} = \alpha^{2}$ and $c_{i} \neq 0$, i.e., $a_{i+1} = 3$ and $a_{i} \neq 0$, or not.

\end{theorem}

\begin{IEEEproof}
We prove Theorem~\ref{thm_rule} directly through the nature of the forbidden pattern.

Consider the symbol $c_i$ of $\bold{c}$. We want to compute the incremental contribution of $c_i$, denoted by $g_i(c_i)$, to the overall index $g(\bold{c})$. This contribution is the number of codewords in $\mathcal{TC}^4_{m}$ starting with $c_{m-1} c_{m-2} \dots c_{i+1}$ from the left and preceding the codeword $c_{m-1} c_{m-2} \dots c_{i+1} c_i \bold{0}^i$ according to the lexicographic ordering \cite{cover_lex}. First of all, notice that:
\begin{itemize}
\item If $c_i = 0$, $g_i(c_i) = 0$.
\item If $c_i \neq 0$, then we have two scenarios. We now elaborate these two scenarios: 
\begin{itemize}
\item First scenario: $c_{i+1} \neq \alpha^{2}$. \\
In this case, the contribution $g_i(c_i)$ equals the number of codewords in $\mathcal{TC}^4_{i}$ (of length $i$) multiplied by $\mathcal{L}(c_i)$, which gives $a_i N(i)$. So, the contribution in this scenario (denoted by $g_{i}'(c_i)$) is:
\begin{align}\label{gi1}
g_{i}'(c_i) = a_i N(i)
\end{align}

\item Second scenario: $c_{i+1} = \alpha^{2}$. \\
In this case, not all codewords in $\mathcal{TC}^4_{i}$ appear on $c_{i-1} c_{i-2} \dots c_0$, which are the $i$ RMSs of $\bold{c}$ in $\mathcal{TC}^4_{m}$. In particular, if $c_i$ were to be  $0$, $c_{i-1}$ equals $\alpha^{2}$ would not be allowed; otherwise, it would produce the forbidden pattern in $\mathcal{Q}^4$ on $c_{i+1} c_i c_{i-1}$. Hence, we should subtract at most $N(i-1)$ from $a_i N(i)$ in order to get the contribution $g_i(c_i)$. However, notice that when we subtract $N(i-1)$, we are also subtracting all sequences of length $m$ where codewords in $\mathcal{TC}^4_{i-2}$ (of length $i-2$) appear on $c_{i-3} c_{i-4} \dots c_0$ such that $c_{i-1}=\alpha^{2}$, $c_{i-2} = 0$, and $c_{i-3} = \alpha^{2}$. While it is correct to exclude these sequences of length $m$ since they include the pattern to be forbidden in $\mathcal{Q}^4$, they are going to be excluded anyways while computing the contribution of $c_{i-2}$ to $g(\bold{c})$, which is $g_{i-2}(c_{i-2})$. In other words, they are taken care of anyways at a smaller length. Thus, we should subtract $N(i-1)$ from $a_i N(i)$, but we should also add at most $N(i-3)$ (the jump is always two indices to the right) in order to prevent over-subtraction. The same procedure is repeated for $N(i-3)$, and so on. In summary, in order to avoid over-subtraction, we adopt the inclusion-exclusion principle to compute $g_i(c_i)$. So, the contribution at this scenario (denoted by $g_{i}''(c_i)$), where $\mathbbm{1}(a_{i+1} = 3 \textup{ \& }   a_{i}\neq 0) = 1$, is:
\begin{align}\label{gi2}
g_{i}''(c_i) = \hspace{-2.7em}\sum_{\hspace{+2.5em}j \geq 0, \textup{ } i-2j > 0} \hspace{-2.3em} (-1)^{j+1}N(i-2j-1)
\end{align}
\end{itemize}
\end{itemize}

Consequently, combining the contribution of each case whether $c_i$ is zero or not and each scenario in (\ref{gi1}) and (\ref{gi2}) for all $m$ symbols of the codeword, the index $g(\bold{c})$ can be written as: 
\begin{align}
g(\bold{c}) &= \sum_{i=0}^{m-1} \Big [ a_i N(i) \nonumber \\ &+ \mathbbm{1}(a_{i+1} = 3 \textup{ \& }   a_{i}\neq 0) \hspace{-2.7em}\sum_{\hspace{+2.5em}j \geq 0, \textup{ } i-2j > 0} \hspace{-2.3em} (-1)^{j+1}N(i-2j-1) \Big ], \nonumber
\end{align}
which completes the proof.
\end{IEEEproof}

\begin{example}\label{example1}

Consider the TD-LOCO codes $\mathcal{TC}^4_{m}$ with $m \in \{2, 3, \dots, 6\}$. We use Theorem~\ref{thm_rule} to compute the index of the TD-LOCO codeword $1\alpha^21\alpha^2\alpha0$ in $\mathcal{TC}^4_{6}$, which is 1824. First, we need to compute the cardinalities needed using Theorem~\ref{thm_card}.
From (\ref{eqn_cardinit}), the defined cardinalities are:
\begin{equation}
N(1) \triangleq 4,  N(0) \triangleq 1, \text{ and } N(-1) \triangleq 1/3. \nonumber
\end{equation}

We compute the required cardinalities as follows:
\begin{align}
N(2) &= 4N_4(1) - N(0) + 3N(-1) = 16, \nonumber \\
N(3) &= 4N_4(2) - N(1) + 3N(0) = 63, \nonumber \\
N(4) &= 4N_4(3) - N(2) + 3N(1) = 248, \nonumber \\
N(5) &= 4N_4(4) - N(3) + 3N(2) = 977, \text{ and} \nonumber \\
N(6) &= 4N_4(5) - N(4) + 3N(3) = 3849. \nonumber
\end{align}

After we get the required cardinalities from Theorem~\ref{thm_card}, we now use Theorem~\ref{thm_rule} to compute the index of the TD-LOCO codeword $1\alpha^21\alpha^2\alpha0$.
\vspace{-0.1em}\begin{align}
g(\bold{c}) &= \sum_{i=0}^{5} \Big [ a_i N(i) \nonumber \\ &+ \mathbbm{1}(a_{i+1} = 3 \textup{ \& }   a_{i}\neq 0) \hspace{-2.7em}\sum_{\hspace{+2.5em}j \geq 0, \textup{ } i-2j > 0} \hspace{-2.3em} (-1)^{j+1}N(i-2j-1) \Big ] \nonumber \\ &= N(5) + 3N(4) + [N(3)+(-N(2)+N(0))] \nonumber \\ &+ 3N(2) + [2N(1)+(-N(0))] \nonumber \\ &= 977 + 3 \times 248 + [63+(-16+1))] \nonumber \\ &+ 3 \times 16 +  [2\times4 +(-1)] = 1824, \nonumber
\end{align}
which is the correct index.
\end{example}

\section{Achievable Rates of Our Coding Scheme}\label{sec_rate}

We first discuss bridging and self-clocking of TD-LOCO codes before calculating the achievable rates of the coding scheme.  
Bridging is needed at the transition between one codeword and another in order to prevent forbidden patterns from arising \cite{ahh_loco}. For instance, given the TD-LOCO code with $m=4$, $\mathcal{TC}^4_{4}$, the forbidden pattern $\alpha^20\alpha^2$ emerges if we write the codeword $01\alpha\alpha^2$ followed by the codeword $0\alpha^21\alpha^2$ on $8$ consecutive grid columns as $01\alpha\underline{\alpha^20\alpha^2}1\alpha^2$ (mapped to binary representation of GF($8$) symbols, see Example~\ref{example2}). Bridging is required to prevent such problems. We perform bridging in a TD-LOCO code $\mathcal{TC}^4_{m}$ by adding a bridging symbol as follows:
\begin{enumerate}
\item If both the RMS of a codeword and the LMS of the next codeword are $\alpha^{2}$'s, bridge with ${\alpha^{2}}$. 
\item Otherwise, bridge with $0$.
\end{enumerate}
If we apply this bridging method to the previous example, we get the stream $01\alpha\underline{\alpha^200\alpha^2}1\alpha^2$, which does not contain the forbidden pattern between codewords after we bridge with $0$. This bridging is both simple and also optimal in terms of protecting codeword edges from ISI and ITI.

\begin{remark}
Note that here we only bridge with one column of $3$ bits vertically. If we were to consider also horizontal bridging on the two-dimensional grid, i.e., bridging with $1$ row of $m$ bits, the redundancy resulting from bridging becomes $3+m$ bits, which grows with $m$, thus notably reducing the rate. Such horizontal bridging is not needed since, as discussed earlier, ITI across groups of $3$ tracks is negligible \cite{chan_tdmr}.
\end{remark}

To maintain calibration of the system, self-clocking is needed \cite{ahh_loco}. Long streams of the same symbol to be written (transmitted) are not allowed in a self-clocked constrained code. If a same-symbol codeword is consecutively transmitted in a stream, as long as this symbol is neither $0$ nor $\alpha^2$, we bridge by a different symbol between each two instances. Thus, a transition always exists across codewords unless the codeword is all $0$'s or all $\alpha^2$'s. Thus, to achieve self-clocking, the two codewords $\bold{0}^m$ or $(\boldsymbol{\alpha^2})^m$ need to be removed from the TD-LOCO code $\mathcal{TC}^4_{m}$.

\begin{definition}\label{def_stdloco}
Let $\mathcal{TC}^4_{m}$ be a TD-LOCO code with $m \geq 1$. A self-clocked TD-LOCO code (CTD-LOCO code) $\mathcal{TC}^{4,\textup{c}}_{m}$ is obtained from $\mathcal{TC}^4_{m}$ as follows:
\begin{equation}\label{eqn_ctdloco}
\mathcal{TC}^{4,\textup{c}}_{m} \triangleq \mathcal{TC}^4_{m}\setminus\{\bold{0}^m,(\boldsymbol{\alpha^2})^m \}.
\end{equation}
Therefore, the cardinality of the CTD-LOCO code is:
\begin{equation}\label{eqn_cTCard}
N^{\textup{c}}(m) = N(m)-2.
\end{equation}
\end{definition}

Define $k^{\textup{c}}_{\textup{eff}}$ as the maximum number of consecutive symbol durations between two consecutive transitions in a stream of CTD-LOCO codewords separated by a bridging symbol. We can also define $k^{\textup{c}}_{\textup{eff}}$ as the maximum number of consecutive grid columns with the exact same $3$ bits after writing to the two-dimensional grid through the CTD-LOCO code. In a way similar to \cite{ahh_qaloco} and \cite{ahh_loco} when $x=1$, $k^{\textup{c}}_{\textup{eff}}$ is given by:
\begin{equation}\label{eqn_keff}
k^{\textup{c}}_{\textup{eff}} = 2(m-1)+1 = 2m-1.
\end{equation}

\begin{table}
\caption{Rates and Normalized Rates of the Overall Coding Scheme Adopting CTD-LOCO Codes $\mathcal{TC}^{4,\textup{c}}_{m}$ (for TDMR Systems)}
\vspace{-0.1em}
\centering
\scalebox{1.1}
{
\begin{tabular}{|c|c|c|}

\hline
{$m$} & $R^{\textup{c}}_{\textup{TD-LOCO}}$ & $R^{\textup{c},\textup{n}}_{\textup{TD-LOCO}}$ \\
\hline
$24$ & $2.8800$ & $0.9600$ \\
\hline
$33$ & $2.9118$ & $0.9706$ \\
\hline
$39$ & $2.9250$ & $0.9750$  \\
\hline
$66$ & $2.9403$ & $0.9801$  \\
\hline
$88$ & $2.9550$ & $0.9850$  \\
\hline
$265$ & $2.9700$ & $0.9900$ \\
\hline
Capacity & $2.9780$ & $0.9927$  \\
\hline
\end{tabular}}
\label{table1}
\end{table}

Now that we have the cardinality of the self-clocked code, we are ready to discuss the achievable rates of our coding scheme adopting TD-LOCO codes. Consider a CTD-LOCO code $\mathcal{TC}^{4,\textup{c}}_{m}$ with cardinality $N^{\textup{c}}(m)$, which is given in (\ref{eqn_cTCard}). The length of the messages $\mathcal{TC}^{4,\textup{c}}_{m}$ encodes in bits is:
\begin{equation}\label{sc}
s^{\textup{c}} = \left \lfloor \log_2 N^{\textup{c}}(m) \right \rfloor = \left \lfloor \log_2 \left( N(m)-2 \right) \right \rfloor.
\end{equation}
Recall that input information message is intentionally selected to be a binary message in order to minimize the number of omitted codewords from $\mathcal{TC}^{4,\textup{c}}_{m}$, and therefore maximize the rate \cite{ahh_qaloco}. The rate of the overall coding scheme adopting the CTD-LOCO code $\mathcal{TC}^{4,\textup{c}}_{m}$ then is:
\begin{equation}\label{eqn_rate}
R^{\textup{c}}_{\textup{TD-LOCO}} = {\frac{s^{\textup{c}}}{m+1}} + 1 = {\frac{\left \lfloor \log_2 \left( N(m)-2 \right) \right \rfloor}{m+1}}+1,
\end{equation}
where $R^{\textup{c}}_{\textup{TD-LOCO}}$ is measured in information bits per coded symbol. Observe that the $+1$ in the denominator comes from the bridging as we always add $1$ symbol for bridging. Meanwhile, the other $+1$ in the equation comes from the additional information bit used for selection while demapping GF($4$) to GF($8$) as discussed in Section~\ref{sec_freebit}. 

Recall from Section~\ref{sec_freebit} that each GF($8$) symbol corresponds to $3$ bits, $2$ bits from the proposed codes in GF($4$) and $1$ additional information bit. Hence, we can normalize the rate in (\ref{eqn_rate}) via dividing by $3$:
\begin{equation}\label{eqn_ratenorm}
R^{\textup{c},\textup{n}}_{\textup{TD-LOCO}} = {\frac{1}{3}}\Bigg({\frac{\left \lfloor \log_2 \left( N(m)-2 \right) \right \rfloor}{m+1}}+1 \Bigg).
\end{equation}

Except only the two codewords $\bold{0}^m$ and $(\boldsymbol{\alpha^2})^{m}$, because of self-clocking as discussed above, all the codewords satisfying the $\mathcal{Q}^4$ constraint are in the CTD-LOCO code $\mathcal{TC}^{4,\textup{c}}_{m}$. We also add $1$ symbol for bridging, and we have 1 additional information bit for selection. Thus, both our CTD-LOCO codes and the overall coding scheme adopting them are \textbf{capacity-achieving}, i.e., the asymptotic rates of the proposed codes and the coding scheme match the respective capacities.

Table~\ref{table1} illustrates the rates and the normalized rates of the overall coding scheme adopting CTD-LOCO codes $\mathcal{TC}^{4,\textup{c}}_{m}$ for various values of $m$. Capacity and normalized capacity of this coding scheme adopting $\mathcal{Q}^4$-constrained codes, which are computed in (\ref{capacity_new}), are also included in the table.

The table demonstrates that the proposed coding scheme with CTD-LOCO codes is capacity-achieving. Even at the smaller lengths $m=24$, the rate is only $3\%$ away from capacity. As $m$ increases, rate increases, and at the moderate length $m=66$, rate is only $1\%$ away from capacity. In addition, the table shows that the proposed coding scheme achieves normalized rates $> 0.97$, i.e., rates $> 0.97 \log_2 8$ information bits per coded symbol. In other words, at acceptable complexities, significant ISI and ITI mitigation is achieved with only $3\%$ or less redundancy.

There are other TD constrained coding schemes, also called bit-stuffing schemes, in the literature. TD-RLL codes with parameters $(d, k)$ satisfy the constraint that each two consecutive $1$'s are separated by at least $d$ and at most $k$ $0$'s both horizontally and vertically \cite{sharov_TCon, halevy_TD, kamabe_TD, kato_TCon, siegel_TCon}. While a TDMR system coded via $(d, k)$ TD-RLL codes will have mitigated ISI and ITI, there are many forbidden sequences that are not error-prone, which results in a notable rate loss. TD constrained codes satisfying the no-isolated-bit (NIB) constraint have higher rates \cite{sharov_TCon, halevy_TD, kamabe_TD} (see Remark~\ref{remark1}). However, these codes do not take into account the observation that ITI across groups of $3$ tracks is negligible if a wide read head is employed in the TDMR system \cite{chan_tdmr}. Thus, for such TDMR systems, our TD-LOCO codes exploit the properties of the system to maximize the rate with a high level of ISI and ITI mitigation.

\section{Algorithms and Writing-Reading}\label{sec_algr}

Here, we introduce the encoding and decoding algorithms of the proposed TD-LOCO codes, which perform the mapping-demapping between an index and the corresponding codeword. The algorithms demonstrated here are based on the encoding-decoding rule (\ref{eqn_newg}) under Theorem~\ref{thm_rule}, and they are a major advantage of the proposed codes since they offer simplicity for the enumerative scheme.

Algorithm~\ref{alg_enc} is the encoding algorithm of our codes. While generating a specific codeword $\bold{c}$ in the algorithm, the RMS of the previous codeword is defined as $\zeta_0$. Note that, when we write on the two-dimensional grid, for each GF($4$) symbol in the output of Algorithm~\ref{alg_enc}, we take $1$ additional information bit. Based on that additional information bit, we decide which of the $2$ GF($8$) symbols mapped into the GF($4$) symbol is the one to be written. Then, this GF($8$) symbol is converted to $3$ binary bits (standard mapping) which are written on a grid column finally. Note that the output of Algorithm~\ref{alg_enc} is not written directly; it is mapped to GF($8$) symbols the binary representation of which is written on the grid. 

\begin{example}\label{example2}
Assume we have a CTD-LOCO codeword of length $m = 5$, which has its symbols in GF($4$). We will write on $5$ different columns in $3$ adjacent tracks, $3$ binary bits each. These $3$ binary bits represent an element in GF($8$). We will write onto the two-dimensional grid as follows:

Assume we have the sequence of incoming input data bits $11101110110110$. In our encoding scheme, it is as if the input data sequence has two parts. The first part is of length $s^{\textup{c}}$, whereas the second part is of length $m$. We know already from (\ref{sc}) for a specific $m$ ($m=5$ in this example) what the message length will be. Here, and using Example~\ref{example1}:
\begin{equation}
s^{\textup{c}} = \left \lfloor \log_2 \left( N(5)-2 \right) \right \rfloor = \left \lfloor \log_2 \left( 977-2 \right) \right \rfloor = 9. \nonumber
\end{equation}

Hence, the first $9$ bits of the input data sequence, underlined in $\underline{111011101}10110$, represent the message, whereas the last $5$~bits, underlined in $111011101\underline{10110}$, represent the additional information bits to select the $5$ GF($8$) symbols for the $5$ columns, respectively.

The message is the input of Algorithm~\ref{alg_enc}. After following the steps of this encoding algorithm, the CTD-LOCO codeword obtained is $1\alpha^2\alpha\alpha^20$. 

Every single time we are going to write on a grid column, we get the respective additional information bit from $\underline{10110}$ to select the GF($8$) symbol. Thus, the de-mapping of the CTD-LOCO codeword $1\alpha^2\alpha\alpha^20$ from GF($4$) to GF($8$) as illustrated in Section~\ref{sec_freebit} results in:  
\begin{center}
\begin{tabular}{lcl}

$1$ & $\underline{1}\longrightarrow$ & $\beta^5$, \\
$\alpha^2$ & $\underline{0}\longrightarrow$ & $0$,\\
$\alpha$ & $\underline{1}\longrightarrow$ & $\beta^3$,\\
$\alpha^2$ & $\underline{1}\longrightarrow$ & $\beta^6$,\\
$0$ & $\underline{0}\longrightarrow$ & $\beta$.\\
\end{tabular}
\end{center}

Finally, for each GF($8$) symbol, its binary representation is written on the $3$ cells of the respective column of the two-dimensional grid:

\begin{center}
\begin{tikzpicture}
\draw[step=0.5cm] (-1,-1) grid (1.5,0.5);

\node at (-0.75,+0.75) {$\beta^5$};
\node at (-0.75,+0.25) {$1$};
\node at (-0.75,-0.25) {$1$};
\node at (-0.75,-0.75) {$0$};

\node at (-0.25,+0.75) {$0$};
\node at (-0.25,+0.25) {$0$};
\node at (-0.25,-0.25) {$0$};
\node at (-0.25,-0.75) {$0$};

\node at (0.25,+0.75) {$\beta^3$};
\node at (0.25,+0.25) {$1$};
\node at (0.25,-0.25) {$0$};
\node at (0.25,-0.75) {$0$};

\node at (0.75,+0.75) {$\beta^6$};
\node at (0.75,+0.25) {$1$};
\node at (0.75,-0.25) {$1$};
\node at (0.75,-0.75) {$1$};

\node at (1.25,+0.75) {$\beta$};
\node at (1.25,+0.25) {$0$};
\node at (1.25,-0.25) {$1$};
\node at (1.25,-0.75) {$0$};
\end{tikzpicture}
\end{center}

This completes the example that shows how to write on the two-dimensional grid given an input data sequence.
\end{example}

\begin{algorithm}\label{alg_enc}
\caption{Encoding CTD-LOCO Codes}
\begin{algorithmic}[1]
\State \textbf{Input:} Incoming binary messages.
\State Use (\ref{eqn_cardgen}) and (\ref{eqn_cardinit}) to compute $N(i)$, $i \in \{2, 3, \dots\}$.
\State Specify $m$, the smallest $i$ in Step~2 to achieve the desired rate. Then, $s^{\textup{c}} = \left \lfloor \log_2 \left( N(m) - 2 \right )  \right \rfloor$.
\State Use (\ref{gi2}) to compute $\textup{inner\_sum}(i)$ for all $i \in \{m-1, m-2, \dots, 0\}$ for the specified $m$.
\State \textbf{for} each incoming message $\bold{b}$ of length $s^{\textup{c}}$ \textbf{do}
\State \hspace{2ex} Compute $g(\bold{c})=\textup{decimal}(\bold{b})+1$.
\State \hspace{2ex} Initialize $\textup{residual}$ with $g(\bold{c})$ and $c_i$ with $0$ for $i \geq m$.
\State \hspace{2ex} \textbf{for} $i \in \{m-1, m-2, \dots, 0\}$ \textbf{do} \textit{(in order)}
\State \hspace{4ex} Set $\textup{index} = i$.
\State \hspace{4ex} \textbf{if} $c_{i+1} = \alpha^2$ \textbf{ then}
\State \hspace{6ex} $\textup{residual\_new} \leftarrow \textup{residual} + \textup{inner\_sum}(\textup{index})$.
\State \hspace{4ex} \textbf{else} \State \hspace{6ex} $\textup{residual\_new} \leftarrow \textup{residual}$.
\State \hspace{4ex} \textbf{end if}
\State \hspace{4ex} \textbf{for} $a_i \in \{1, 2, 3\}$ \textbf{do}
\State \hspace{6ex} \textbf{if} $(a_i-1)N(\textup{index}) \leq \textup{residual\_new} < a_i N(\textup{index})$ \textbf{then}
\State \hspace{8ex} Encode $c_i = \mathcal{L}^{-1}(a_i-1)$.
\State \hspace{8ex} $\textup{residual} \leftarrow \textup{residual} - (a_i-1) N(\textup{index})$.
\State \hspace{8ex} \textbf{break}. \textit{(exit current loop)}
\State \hspace{6ex} \textbf{end if}
\State \hspace{4ex} \textbf{end for}
\State \hspace{4ex} \textbf{if} $\textup{residual\_new} \geq 3 N(\textup{index})$ \textbf{then}
\State \hspace{6ex} Encode $c_i = \alpha^2$. 
\State \hspace{6ex} $\textup{residual} \leftarrow \textup{residual} - 3N(\textup{index})$.
\State \hspace{4ex} \textbf{end if}
\State \hspace{4ex} \textbf{if} (not first codeword) $\land$ ($i = m-1$) \textbf{then}
\State \hspace{6ex} \textbf{if} ($\zeta_0 = \alpha^{2}$) $\land$ ($c_{m-1} = \alpha^{2}$) \textbf{then}
\State \hspace{8ex} Bridge with ${\alpha^{2}}$ before $c_{m-1}$.
\State \hspace{6ex} \textbf{else}
\State \hspace{8ex} Bridge with $0$ before $c_{m-1}$.
\State \hspace{6ex} \textbf{end if}
\State \hspace{4ex} \textbf{end if}
\State \hspace{2ex} \textbf{end for}
\State \textbf{end for}
\State \textbf{Output:} Outgoing stream of CTD-LOCO codewords. \textit{(binary representation of GF($8$) symbols is to be written on the grid columns)}
\end{algorithmic}
\label{alg_enc}
\end{algorithm}

Algorithm~\ref{alg_dec} is the decoding algorithm of our codes, and it is a direct implementation of Theorem~\ref{thm_rule}.

Reversing the procedure in Example~\ref{example2} serves to illustrate the reading process.

\begin{algorithm}
\caption{Decoding CTD-LOCO Codes}
\begin{algorithmic}[1]
\State \textbf{Inputs:} Incoming stream of  CTD-LOCO codewords, in addition to $m$, and $s^{\textup{c}}$.
\State Use (\ref{eqn_cardgen}) and (\ref{eqn_cardinit}) to compute $N(i)$, $i \in \{2, 3, \dots, m\}$.
\State Use (\ref{gi2}) to compute $\textup{inner\_sum}(i)$ for all $i \in \{m-1, m-2, \dots, 0\}$ for the given $m$.
\State \textbf{for} each incoming codeword $\bold{c}$ of length $m$ \textbf{do}
\State \hspace{2ex} Initialize $g(\bold{c})$ with $0$ and $c_i$ with $0$ for $i \geq m$.
\State \hspace{2ex} \textbf{for} $i \in \{m-1, m-2, \dots, 0\}$ \textbf{do} \textit{(in order)}
\State \hspace{4ex} Set $\textup{index} = i$.
\State \hspace{4ex} \textbf{if} $c_i \neq 0$ \textbf{then} \textit{(same as $a_i \neq 0$)}
\State \hspace{6ex} Set $a_i = \mathcal{L}(c_i)$.
\State \hspace{6ex} \textbf{if} $c_{i+1} = 3$ \textbf{then} \textit{(same as $a_{i+1} = 3$)}
\State \hspace{8ex} $g(\bold{c}) \leftarrow g(\bold{c}) + a_i N(\textup{index}) + \textup{inner\_sum}(\textup{index})$.
\State \hspace{6ex} \textbf{else}
\State \hspace{8ex} $g(\bold{c}) \leftarrow g(\bold{c}) + a_i N(\textup{index})$.
\State \hspace{6ex} \textbf{end if}
\State \hspace{4ex} \textbf{end if}
\State \hspace{2ex} \textbf{end for}
\State \hspace{2ex} Compute $\bold{b}=\textup{binary}(g(\bold{c})-1)$, which has length $s^{\textup{c}}$.
\State \hspace{2ex} Ignore the next $1$ bridging symbol.
\State \textbf{end for}
\State \textbf{Output:} Outgoing binary messages.
\end{algorithmic}
\label{alg_dec}
\end{algorithm}

\begin{table}[H]
\caption{Complexity (adder size) with Respect to Length of CTD-LOCO Codes, $\mathcal{TC}^{4,\textup{c}}_{m}$}
\vspace{-0.1em}
\centering
\scalebox{1.1}
{
\begin{tabular}{|c|c|}

\hline
{$m$} & Complexity \\
\hline
$24$ & $47$ \\
\hline
$33$ & $65$ \\
\hline
$39$ & $77$ \\
\hline
$66$ & $130$ \\
\hline
$88$ & $174$ \\
\hline
\end{tabular}}
\label{table2}
\end{table}

As observed from the algorithms above, encoding and decoding of TD-LOCO codes are performed through simple adders. Adders perform comparisons, subtractions, and additions specified in the encoding and decoding algorithms, and the size of an adder is the message length $s^\textup{c}$, which is calculated in (\ref{sc}). Table~\ref{table2} shows how the complexity (adder size) changes with the length of the CTD-LOCO code. Observe that the adder size can also be computed by multiplying rate of the CTD-LOCO code at any length $m$ by $m+1$ (the rate here is not the rate of the coding scheme adopting the CTD-LOCO code illustrated in Table~\ref{table1}; it is rather the rate of the CTD-LOCO code itself). Table~\ref{table1} demonstrates that a rate of $0.97$ is achievable with an adder size of $65$ bits, which is a modest complexity (see the entry of $m=33$).

TD-LOCO codes are reconfigurable. When the adder sizes are appropriate, same set of adders used to encode-decode a specific TD-LOCO code can be reconfigured by changing their inputs (cardinalities) through simple multiplexers to encode-decode another TD-LOCO code with a different constraint if applicable. Forbidding the SIS pattern can be seen as the constraint to support when the device is fresh, i.e., early in the lifetime of the magnetic disk. As the device ages, and thus deteriorates, new error-prone patterns to constrain arise. Reconfigurability of TD-LOCO codes will be useful to reconfigure our codes to support new constraints and forbid these new patterns (if families of TD-LOCO codes can be constructed for such constraints, which is among our future work). More details about reconfigurability can be found in \cite{ahh_qaloco} and \cite{ahh_loco}.

A promising future research direction is to combine efficient TD constrained codes with high performance multi-dimensional low-density parity-check (LDPC) codes \cite{ahh_mdg} and investigate the performance gains in TDMR systems.

\section{Conclusion}\label{sec_conc}

We have proposed constrained codes, TD-LOCO codes, for TDMR storage systems for which we focused on wide read heads that read three tracks simultaneously. With the observation of error-prone nature of SIS patterns in the recent literature, we designed constrained codes for read channels where SIS patterns are highly vulnerable to error. The central idea is to translate the problem of designing constrained codes for binary two-dimensional constraint to that of designing codes for a non-binary one-dimensional constraint. After introducing the base case TD-LOCO codes with this mapping, we simplified the problem of non-binary code design by organizing GF($8$) symbols (binary triplets) into four pairs, and constructing constrained codes over GF($4$). This reduces complexity because additional bit that selects the representative from the pair is uncoded - it does not enter into the algorithms for encoding and decoding the constrained code. We showed that the shift to constrained coding over GF($4$), mapping with additional information bit idea, incurs a negligible loss in capacity (only $0.54\%$ away from the optimal capacity). Following this new scheme, we proposed our novel TD-LOCO codes, provided their cardinality, and derived their encoding-decoding rule. We have described simple algorithms for encoding and decoding, as well as bridging and self-clocking, and we have discussed reconfigurability. 
To sum up, the aim of TDMR technology is to achieve storage densities that are competitive with Flash memory. The constrained codes presented here are designed to improve reliability of the recent TDMR technology. With the proposed coding scheme adopting TD-LOCO codes, the redundancy required for device protection by the constrained code is less than $3\%$ with acceptable complexity, which preserves the density gains achieved by TDMR devices even after we combine with error-correcting codes.

\section*{Acknowledgment}

This research was supported in part by NSF under Grant CCF 1717602 and in part by AFOSR under Grant FA 9550-17-1-0291.

\ifCLASSOPTIONcaptionsoff
  \newpage
\fi



\begin{thebibliography}{18}

\balance

\bibitem{veeresh_mlc}
V. Taranalli, H. Uchikawa, and P. H. Siegel, ``Error analysis and inter-cell interference mitigation in multi-level cell flash memories,'' in \emph{Proc. IEEE Int. Conf. Commun. (ICC)}, London, UK, Jun. 2015, pp. 271--276.

\bibitem{ahh_qaloco}
A. Hareedy, B. Dabak, and R. Calderbank, ``Managing device lifecycle: reconfigurable constrained codes for M/T/Q/P-LC Flash memories,'' Jan. 2020. [Online]. Available: https://arxiv.org/abs/2001.02325

\bibitem{wood_tdmr}
R. Wood, M. Williams, A. Kavcic, and J. Miles, ``The feasibility of magnetic recording at 10 terabits per square inch on conventional media,'' \emph{IEEE Trans. Magn.}, vol. 45, no. 2, pp. 917--923, Feb. 2009.

\bibitem{chan_tdmr}
K. S. Chan and M. R. Elidrissi, ``A system level study of two-dimensional magnetic recording (TDMR),'' \emph{IEEE Trans. Magn.}, vol. 49, no. 6, pp. 2812--2817, Jun. 2013.

\bibitem{shayan_tdmr}
S. G. Srinivasa, Y. Chen, and S. Dahandeh, ``A communication-theoretic framework for 2-DMR channel modeling: performance evaluation of coding and signal processing methods,'' \emph{IEEE Trans. Magn.}, vol. 50, no. 3, pp. 6--12, Mar. 2014.

\bibitem{mohsen_tdmr}
M. Bahrami, C. K. Matcha, S. M. Khatami, S. Roy, S. G. Srinivasa, and B. Vasic, ``Investigation into harmful patterns over multitrack shingled magnetic detection using the Voronoi model,'' \emph{IEEE Trans. Magn.}, vol. 51, no. 12, pp. 1--7, Dec. 2015.

\bibitem{pituso_tdmr}
K. Pituso, C. Warisarn, D. Tongsomporn, and P. Kovintavewat, ``An intertrack interference subtraction scheme for a rate-4/5 modulation code for two-dimensional magnetic recording,'' \emph{IEEE Magn. Letters}, vol. 7, pp. 1--5, Jul. 2016.

\bibitem{victora_10tbpsi}
R. H. Victora, S. M. Morgan, K. Momsen, E. Cho, and M. F. Erden, ``Two-dimensional magnetic recording at 10 tbpsi,''
\emph{IEEE Trans. Magn.}, vol. 48, no. 5, pp. 1697--1703, May 2012.

\bibitem{hwang_10tb}
E. Hwang, R. Negi, and B. V. K. Kumar, ``Signal processing for near 10 Tbpsi density in two-dimensional magnetic recording (TDMR),'' \emph{IEEE Trans. Magn.}, vol. 46, no. 6, pp. 1813--1816, Jun. 2010.

\bibitem{seagate}
M. Re, ``Tech talk on HDD areal density,'' \emph{Seagate}, Aug. 2015. [Online]. Available: \href{https://web.archive.org/web/20180528133250/https://www.seagate.com/www-content/investors/\_shared/docs/tech-talk-mark-re-20150825.pdf}{hdd\_areal\_density\_seagate}

\bibitem{mallary_1tb}
M. Mallary, A. Torabi and M. Benakli, ``One terabit per square inch perpendicular recording conceptual design,'' \emph{IEEE Trans. Magn.}, vol. 38, no. 4, pp. 1719--1724, July 2002.

\bibitem{tang_bahl}
D. T. Tang and R. L. Bahl, ``Block codes for a class of constrained noiseless channels,'' \emph{Inf. and Control}, vol. 17, no. 5, pp. 436--461, 1970.

\bibitem{siegel_mr}
P. Siegel, ``Recording codes for digital magnetic storage,'' \emph{IEEE Trans. Magn.}, vol. 21, no. 5, pp. 1344--1349, Sep. 1985.

\bibitem{ach_fsm}
R. Adler, D. Coppersmith, and M. Hassner, ``Algorithms for sliding block codes--An application of symbolic dynamics to information theory,'' \emph{IEEE Trans. Inf. Theory}, vol. 29, no. 1, pp. 5--22, Jan. 1983.

\bibitem{immink_surv}
K. A. S. Immink, P. H. Siegel, and J. K. Wolf, ``Codes for digital recorders,'' \emph{IEEE Trans. Inf. Theory}, vol. 44, no. 6, pp. 2260--2299, Oct. 1998.

\bibitem{ahh_loco}
A. Hareedy and R. Calderbank, ``LOCO codes: lexicographically-ordered constrained codes,''  \emph{IEEE Trans. Inf. Theory}, vol. 66, no. 6, pp. 3572--3589, Jun. 2020.

\bibitem{ahh_aloco}
A. Hareedy and R. Calderbank, ``Asymmetric LOCO codes: constrained codes for Flash memories,'' in \emph{Proc. 57th Annual Allerton Conf. Commun., Control, and Computing}, Monticello, IL, USA, Sep. 2019, pp. 124--131.

\bibitem{sharov_TCon}
A. Sharov and R. M. Roth, ``Two-Dimensional Constrained Coding Based on Tiling,'' \emph{IEEE Trans. Inf. Theory}, vol. 56, no. 4, pp. 1800--1807, Apr. 2010.

\bibitem{halevy_TD}
S. Halevy, J. Chen, R. M. Roth, P. H. Siegel, and J. K. Wolf, ``Improved bit-stuffing bounds on two-dimensional constraints,'' \emph{IEEE Trans. Inf. Theory}, vol. 50, no. 5, pp. 824--838, May 2004.

\bibitem{kamabe_TD}
H. Kamabe, ``Constraints for two-dimensional recording media,'' in \emph{Proc. IEEE Int. Symp. Inf. Theory Appl. (ISITA)}, Auckland, New Zealand, Dec. 2008, pp. 1--6.

\bibitem{kato_TCon}
A. Kato and K. Zeger, ``On the capacity of two-dimensional run-length constrained channels,'' \emph{IEEE Trans. Inf. Theory}, vol. 45, no. 5, pp. 1527--1540, Jul. 1999.

\bibitem{siegel_TCon}
P. H. Siegel and J. K. Wolf, ``Bit-stuffing bounds on the capacity of 2-dimensional constrained arrays,'' in \emph{Proc. IEEE Int. Symp. Inf. Theory (ISIT)}, Cambridge, MA, USA, Aug. 1998, pp. 323.

\bibitem{shannon}
C. E. Shannon, ``A mathematical theory of communication,'' {Bell Sys. Tech. J.}, vol. 27, Oct. 1948.

\bibitem{cover_lex}
T. Cover, ``Enumerative source encoding,'' \emph{IEEE Trans. Inf. Theory}, vol. 19, no. 1, pp. 73--77, Jan. 1973.

\bibitem{ahh_mdg}
A. Hareedy, R. Kuditipudi, and R. Calderbank, ``Minimizing the number of detrimental objects in multi-dimensional graph-based codes,'' \emph{IEEE Trans. Commun.}, to be published, doi: 10.1109/TCOMM.2020.2991072.

\end{thebibliography}
\end{document}